\documentclass[12pt]{article}
\usepackage{epsfig}

\makeatletter
\renewcommand\section{\@startsection {section}{1}{\z@}%
                                   {-3.5ex \@plus -1ex \@minus -.2ex}
                                   {2.3ex \@plus.2ex}%
                                   {\normalfont\large\bfseries}}
\renewcommand\subsection{\@startsection{subsection}{2}{\z@}%
                                     {-3.25ex\@plus -1ex \@minus -.2ex}%
                                     {1.5ex \@plus .2ex}%
                                     {\normalfont\bfseries}}
\makeatother

\def\IZ{\relax\ifmmode\mathchoice
{\hbox{\cmss Z\kern-.4em Z}}{\hbox{\cmss Z\kern-.4em Z}}
{\lower.9pt\hbox{\cmsss Z\kern-.4em Z}} {\lower1.2pt\hbox{\cmsss
Z\kern-.4em Z}}\else{\cmss Z\kern-.4em Z}\fi}
\def\IR{\relax{\rm I\kern-.18em R}}

\def\one{{\hbox{ 1\kern-.8mm l}}}

\newlength{\bredde}
\def\slash#1{\settowidth{\bredde}{$#1$}\ifmmode\,\raisebox{.15ex}{/}
\hspace*{-\bredde} #1\else$\,\raisebox{.15ex}{/}\hspace*{-\bredde}
#1$\fi}

\newsavebox{\zzzbar}
\sbox{\zzzbar}
  {\setlength{\unitlength}{0.9em}
  \begin{picture}(0.6,0.7)
  \thinlines
  \put(0,0){\line(1,0){0.6}}
  \put(0,0.75){\line(1,0){0.575}}
  \multiput(0,0)(0.0125,0.025){30}{\rule{0.3pt}{0.3pt}}
  \multiput(0.2,0)(0.0125,0.025){30}{\rule{0.3pt}{0.3pt}}
  \put(0,0.75){\line(0,-1){0.15}}
  \put(0.015,0.75){\line(0,-1){0.1}}
  \put(0.03,0.75){\line(0,-1){0.075}}
  \put(0.045,0.75){\line(0,-1){0.05}}
  \put(0.05,0.75){\line(0,-1){0.025}}
  \put(0.6,0){\line(0,1){0.15}}
  \put(0.585,0){\line(0,1){0.1}}
  \put(0.57,0){\line(0,1){0.075}}
  \put(0.555,0){\line(0,1){0.05}}
  \put(0.55,0){\line(0,1){0.025}}
  \end{picture}}

\newcommand{\ena}{\end{eqnarray}}
\newcommand{\beqa}{\begin{eqnarray}}
\newcommand{\eeqa}{\end{eqnarray}}
\newcommand{\bea}{\begin{eqnarray}}
\newcommand{\eea}{\end{eqnarray}}

\newcommand{\eq}[1]{(\ref{#1})}

\newcommand{\be}{\begin{equation}}
\newcommand{\ee}{\end{equation}}

\usepackage{graphicx}

\newcommand{\beq}{\begin{equation}}
\newcommand{\eeq}{\end{equation}}
\newcommand{\ber}{\begin{array}}
\newcommand{\eer}{\end{array}}

\newcommand{\del}{\partial}

\newcommand{\dsty}{\displaystyle}

\newcommand{\eps}{\varepsilon}

\begin{document}
\begin{titlepage}
\begin{flushright}
arXiv:0806.3057 [hep-th]
\end{flushright}
\vfill
\begin{center}
{\LARGE\bf On discrete features of the wave equation\vspace{2mm}\\
in singular pp-wave backgrounds}    \\
\vskip 10mm
{\large Oleg Evnin$^a$ and Timothy Nguyen$^b$}
\vskip 7mm
{\em $^a$ Theoretische Natuurkunde, Vrije Universiteit Brussel and\\
The International Solvay Institutes\\ Pleinlaan 2, B-1050 Brussels, Belgium}
\vskip 3mm
{\em $^b$ Massachusetts Institute of Technology \\ Department of Mathematics 2-310 \\ 77 Massachusetts Ave \\ Cambridge, MA 02139}
\vskip 3mm
{\small\noindent  {\tt eoe@tena4.vub.ac.be, timothyn@math.mit.edu}}\end{center}
\vfill

\begin{center}
{\bf ABSTRACT}\vspace{3mm}
\end{center}

We analyze the wave equation in families of pp-wave geometries developing strong
localized scale-invariant singularities in certain limits. For both cases
of well-localized pp-waves and the so-called null-cosmologies, we observe
an intriguing discrete dependence of the existence of a singular limit
on the normalization of the pp-wave profile. We also find restrictive matching
conditions relating the geometries before and after the singularity
(if a singular limit for the solutions of the wave equation with initial
conditions specified away from the near-singular region is assumed to exist).

\vfill

\end{titlepage}

\section{pp-waves}

pp-wave geometries are an interesting family of essentially non-linear solutions of general relativity containing arbitrary functions
of space-time coordinates and describing propagation of strong gravitational waves in space-time. Their importance extends beyond
classical general relativity, since, in a large class of quantum gravitational theories (including perturbative string theories), they
turn out to be unaffected by quantum corrections
(see \cite{horowitzsteif} and references therein). Furthermore, the presence of a light-like Killing vector in these backgrounds
permits both a formulation of the matrix theory and an analytic solution of string theory sigma-models
in the light-cone gauge.

It is interesting to consider the limit whereby the arbitrary functions contained
in the pp-wave solutions develop isolated singularities. The corresponding light-like
singularities in space-time geometry are reminiscent (even though different)
from (space-like) cosmological singularities. Furthermore, the special
standing of pp-waves in quantum gravitational theories makes these geometries
a fruitful playground for studying quantum gravity in extreme high-curvature regime,
and, in particular, exploring the possibilities to define dynamical transitions
through space-time singularities.

In this note, we shall concentrate on the following class of simple pp-wave geometries:
\be
ds^2= - 2dX^+\,dX^- - \lambda F(X^+,\eps)X^2(dX^+)^2+dX^2,
\label{metric}
\ee
where $F(X^+,\eps)$ is an arbitrary function developing a singularity at $X^+=0$ when $\eps$ is sent to 0 and $\lambda$ is a number (the
overall pp-wave profile normalization). We shall formally work in three space-time dimensions, but our derivations can be immediately extended to higher dimensions
by replacing $dX^2$ with $dX^idX^i$.  We shall assume that $F$ does not depend on any dimensional parameters\footnote{Note that even though all our
considerations will, strictly
speaking, apply only to this ``scale-invariant'' case, one should expect that
even if $F$ depends on other dimensional parameters, at least in some cases,
they should not affect the existence of the limit. Indeed, the singularity
transition takes place in an $\eps$-neighborhood of $X^+=0$ and it should not
be particularly sensitive to dimensional parameters that stay finite as
$\eps$ is taken to 0.}
besides $\eps$, in which case the $\eps\to 0$ limit is scale-invariant, and
on dimensional grounds, one can write $F$ as
\be
F(X^+,\eps)=\frac1{\eps^2}\,\Omega(X^+/\eps).
\label{F}
\ee
The case of $F(X^+,\eps)\sim (1/X^+)^2$ has been previously studied\footnote{\label{penrose} Such scale-invariant $(1/X^+)^2$ dependencies of the pp-wave profile arise as Penrose limits of a broad class \cite{limit} of power-law space-time singularities (including the common singularities encountered in cosmology). This observation suggests (though not in a conclusive way) that it is natural
to resolve such singularities in the scale-invariant fashion of (\ref{F}).} in \cite{PRT}. However,
our approach will be quite different, as we shall be considering the singular case
as a limit of regularized geometries (rather than devising singularity transition recipes
for the singular case itself). The motivation for this approach is that if the background
(\ref{metric}) is used in the context of perturbative string theory (and related approaches
to quantum gravity), it must satisfy Einstein's equations (when those are exact, or else an appropriate generalization thereof), and this can only be ensured
by working with a regularized space and taking an $\eps\to 0$ limit in the end. Remarkably,
it is precisely this limiting procedure that will be responsible for the emergence of
discrete structures we are aiming to report.

We intend to study the wave equation for a free scalar field in the background (\ref{metric}).
Whereas the wave equation itself does not describe the physical evolution on strongly
curved spaces adequately, its solutions, the mode functions, are essential ingredients
of any field-theoretical or string-theoretical set-up formulated in the background (\ref{metric}). Likewise, we shall not discuss how
the scalar field should be coupled
to the dilaton or $p$-form potentials necessary to make the metric (\ref{metric}) satisfy Einstein's equations \cite{PRT}, and we shall
assume that the scalar field is minimally coupled to the metric. One can hope that the robustness and genericity of the features
we observe will make them survive at least for some modifications and
extensions\footnote{One natural problem to analyze is the evolution of a free
quantum string in such regularized pp-waves and its singular limit \cite{prep} (see, e.g., \cite{PRT} for related earlier work).}
of our present set-up. Our main observation will be that the $\eps\to 0$ limit of the solutions to the wave equation in the background
(\ref{metric}) with initial data specified away from the singularity typically exist
only for discrete\footnote{This discreteness has been observed for a special class of pp-waves in \cite{g-res}, and our present
objective is to show its generic nature.} values of the parameter $\lambda$ and only if the leading asymptotics of $\Omega$
in (\ref{F}) are the same for $X^+$ going to $+\infty$ and $-\infty$.

\section{The wave equation}

In the background (\ref{metric}), the Klein-Gordon wave equation takes the form
\beq
\del_+\del_-\phi-\frac12\partial^2_X\phi-\frac\lambda2\, F(X^+,\eps)X^2\,\del_-^2\phi+ \frac{m^2}{2}\phi=0,
\label{waeq1}
\eeq
or, after Fourier-transforming,
\begin{equation}
\phi(X^+,X^-,X)=\frac{1}{\sqrt{2 \pi}} \int dk_- \phi_{k_-}(X)\,\mathrm{exp}(i k_- X^-),
\label{fourierTransf}
\end{equation}
it can be re-written in the form
\be
-i\dot{\phi}=-\frac{\partial^2_X\phi}{2 k_-}+\frac{\lambda k_-}2\, F(X^+,\eps)X^2 \phi+ \frac{m^2}{2 k_-}\phi
\label{waeq2}
\ee
(where the dot denotes the $X^+$-derivative, and we have suppressed the $k_-$ index on $\phi$). The latter
representation makes it manifest that the wave equation in pp-wave backgrounds takes the form of a Schr\"odinger equation (a well known
fact, see for example \cite{PRT}).

A general overview of singular limits in time-dependent Schr\"odinger equations, such as the $\eps\to 0$ limit in (\ref{waeq2}), has
been given in section 2 of \cite{g-res}.
It has been noted in particular that, if the Schr\"odinger equation possesses a finite-dimensional {\it dynamical group}, it reduces to
a finite number of ordinary
differential equations, which considerably simplifies the analysis of the singular limit.
The equation (\ref{waeq2}) presents a particularly straightforward realization of
this structure, since it is a Schr\"odinger equation for a linear quantum system,
and as such, it can be reduced to ordinary differential equations (``classical equations of motion'') using the standard WKB
techniques.

More specifically, one proceeds as follows. The formal ``Hamiltonian''
corresponding to (\ref{waeq2}) is
\begin{equation}
\mathcal{H}=\frac{P^2}{2 k_-}+\frac{\lambda k_-}2\, F(t,\eps)X^2 + \frac{m^2}{2 k_-}
\label{hamilt}
\end{equation}
(where $X^+$ has been renamed to $t$ in order to make the ``quantum-mechanical''
considerations look more familiar). Note that (\ref{hamilt}) is nothing but
the Hamiltonian of a harmonic oscillator with a time-dependent frequency.
The ``Schr\"odinger'' equation (\ref{waeq2}) is then solved by the ansatz
\begin{equation}
\phi(X_1,t_1|X_2,t_2)={\cal A}(t_1,t_2)\,\mathrm{exp}\left(-i S_{cl}\left[X_1,t_1|X_2,t_2\right]\right),
\label{WKBansatz}
\end{equation}
where $t_2$ should be identified with $X^+$ of (\ref{waeq2}), if
\bea
S_{cl}=\int_{t_1}^{t_2}\,\mathrm{d}t\,\left(P \dot{X} - \mathcal{H}\right){\Big|_{X=X_{cl}(X_1,t_1|X_2,t_2)}}\label{WKBactie}\\
-2 k_-\,\frac{\partial {\cal A}(t_1,t)}{\partial t}={\cal A}(t_1,t)\,\frac{\partial^2 S_{cl}\left[X_1,t_1|X,t\right]}{\partial
X^2}\label{WKB_pref}
\eea
Here, $S_{cl}$ is the ``classical action'' for the solution satisfying $X(t_1)=X_1,\ X(t_2)=X_2$. More general solutions to \eq{waeq2}
are obtained by integrating \eq{WKBansatz} over $X_1$, weighted by an arbitrary smooth wavepacket.

The classical equation of motion corresponding to (\ref{hamilt}) is
\be
\ddot{X}+\lambda F(t,\eps) X=0
\label{cls}
\ee
Given two independent solutions to this equation, $f(t)$ and $h(t)$, one
can straightforwardly construct the solution $X_{cl}(X_1,t_1|X_2,t_2)$ satisfying
$X(t_1)=X_1,\ X(t_2)=X_2$, and subsequently evaluate the action (\ref{WKBactie}):
\be
S_{cl}=-\frac{k_-}{2}\,\frac{h_2\dot{f}_1-f_2\dot{h}_1}{f_1h_2-h_1f_2}\,X^2_1+
\frac{k_-}{2}\,\frac{f_1\dot{h}_2-h_1\dot{f}_2}{f_1h_2-h_1f_2}\,X^2_2 -k_- \frac{W[f,h]}{f_1h_2-h_1f_2}\,X_1X_2- \frac{m^2}{2
k_-}(t_2-t_1),
\label{scl}
\ee
where $W[f,h]=f\dot{h}-h\dot{f}$ is the Wronskian of $f(t)$ and $h(t)$ (independent of $t$), and we have introduced $f_1=f(t_1),
h_2=h(t_2)$, etc.

With the above form of $S_{cl}$, (\ref{WKB_pref}) reduces to
\begin{equation}\label{pref_general}
\frac{\partial {\cal A}(t_1,t)}{\partial t}=-\frac12\frac{f_1\dot{h}-h_1\dot{f}}{f_1h-h_1f}\,{\cal A}(t_1,t),
\end{equation}
which can be solved as
\begin{equation}\label{pref_soln}
{\cal A}(t_1,t_2)=\frac{\cal N}{\sqrt{f_1h_2-h_1f_2}},
\end{equation}
with a normalization constant $\cal N$. Depending on how the solutions to the wave equation
are to be used, different normalizations can be chosen. This ambiguity will not be relevant
for our considerations, and it will be convenient to think of $\cal N$ as being proportional to $\sqrt{W[f,h]}$, which makes
independence on the normalization of
$f$ and $h$ manifest\footnote{The choice of the branch structure of the square root in (\ref{pref_soln}) is somewhat subtle but
completely unambiguous and is given by the so called {\it Maslov phase} prescription. We shall refer the reader to appendix B of
\cite{g-res} for further details, which will not be relevant as far as our present goals are concerned.}. We thus arrive at the
following complete basis of solutions
(labelled by $X_1$) to the wave equation (\ref{waeq2}):
\begin{equation}
\phi\sim \frac 1{\sqrt{{\cal C}(t_1,t_2)}}\,\exp\left(\frac{ik_-\del_{t_1}{\cal C}}{2{\cal C}}\,X^2_1- \frac{ik_-\del_{t_2}{\cal
C}}{2{\cal C}}\,X^2_2 +\frac{ik_-}{\cal C}\,X_1X_2+ \frac{im^2}{2 k_-}(t_2-t_1)\right),
\label{soln}
\end{equation}
where we have introduced the ``compression factor''
\be
{\cal C}(t_1,t_2)=\frac{f(t_1)h(t_2)-h(t_1)f(t_2)}{W[f,h]}.
\label{comp}
\ee
Note that ${\cal C}(t_1,t_2)$ depends only on equation (\ref{cls}) and not on the choice of solutions $f$ and
$h$. Indeed, if one changes to a different solution basis
\be
\left(\begin{array}{c}\tilde f\vspace{1mm}\\ \tilde h\end{array}\right) = A \left(\begin{array}{c}f\vspace{1mm}\\ h\end{array}\right),
\ee
both numerator and denominator of (\ref{soln}) are multiplied by $\det A$. This property of ${\cal C}(t_1,t_2)$ makes it obvious that
(\ref{soln}) is independent
of the choice of $f$ and $h$, as it should be. (Zeros of ${\cal C}(t_1,t_2)$
correspond to focal points of the equation (\ref{cls}), see appendix B of \cite{g-res}
for further details.) As a matter of fact, ${\cal C}(t_1,t_2)$ can be recognized
as a solution (with respect to $t_2$) to (\ref{cls}) satisfying initial conditions
\be
{\cal C}(t_1,t_2)\Big|_{t_2=t_1}=0,\qquad \del_{t_2}{\cal C}(t_1,t_2)\Big|_{t_2=t_1}=1.
\label{icC}
\ee

\section{The singular limit}

The representation (\ref{soln}) for a complete basis of solutions to the wave
equation (\ref{waeq1}) derived in the previous section is relevant for
our present goals inasmuch as it relates our original problem to a
simple ordinary differential equation (\ref{cls}). In particular,
the existence of an $\eps\to 0$ limit\footnote{We are speaking of an $\eps\to 0$ limit corresponding to a meaningful dynamical evolution across the singularity. The typical situation is that ${\cal C}(t_1,t_2)$ blows up for $t_1<0$, $t_2>0$ in the $\eps\to 0$ limit, i.~e., the harmonic oscillator (\ref{cls}) is knocked
out to infinity by the singular potential. In this case, the $\eps\to 0$ limit of (\ref{soln}) exists in the mathematical sense, and is $\phi(t_2)=0$ for $t_2>0$. This limit is, of course, completely meaningless as far as defining dynamical evolution across the singularity is concerned, and we refer to this situation as having no $\eps\to 0$ limit.} to (\ref{waeq1}) can be completely
analyzed in terms of a single solution\footnote{Note however, that if the $\eps\to 0$ limit of (\ref{icC}) exists for all $t_1$, the
limit will exist for solutions
satisfying arbitrary initial conditions. This follows from the fact that $\del_{t_1}{\cal C}(t_1,t_2)$ satisfies initial conditions
linearly independent of (\ref{icC}). The converse is also obviously true.} (\ref{icC}) to (\ref{cls}).
In this section, we shall analyze the $\eps\to 0$ limit of ${\cal C}(t_1,t_2)$ for
the special (``scale-invariant'') case of (\ref{F}). We shall observe that the limit
exists only for discrete values of the overall normalization of the pp-wave profile, and only if the leading asymptotics of $\Omega$ of (\ref{F}) is the same for its
argument going to plus and minus infinity.

\subsection{Scaling}

For the particular pp-wave given by (\ref{metric}-\ref{F}), the ``classical equation of motion'' (\ref{cls}) takes the form
\be
\ddot{X}+\frac\lambda{\eps^2}\,\Omega(t/\eps) X=0,
\label{clsscl}
\ee
and, as explained above, we are interested in the limiting behavior of the
solution satisfying
\be
X(t_1)=0,\qquad \dot X(t_1)=1
\ee
as $\eps$ is taken to 0.

The scaling properties of (\ref{F}) permit rewriting this equation in a dimensionless
form with $\eta=t/\eps$, $Y(\eta)=X(\eps\eta)$:
\be
Y''+\lambda \Omega(\eta) Y=0, \qquad Y\big|_{\eta=t_1/\eps}=0,\qquad Y'\big|_{\eta=t_1/\eps}=\eps,
\label{Y}
\ee
and, in this representation, one should be looking for an $\eps\to 0$ limit
of $Y(t/\eps)$.

Note that the differential equation (\ref{Y}) is itself $\eps$-independent,
and the $\eps\to 0$ limit has been translated into specifying initial conditions
in the infinite past (we are assuming $t_1$ to be negative), while ``observing''
the results of the evolution at the point $t/\eps$ in the infinite future.
In other words, the requirement that an $\eps\to 0$ limit should exist
has been translated into some constraints on the asymptotic behavior
of solutions to an $\eps$-independent differential equation. This latter
formulation has a conspicuous flavor of a Sturm-Liouville problem,
which makes the appearance of a discrete spectrum for $\lambda$ hardly
surprising. We shall see how this works out explicitly, after considering
the asymptotic behavior of solutions to (\ref{Y}).

\subsection{Asymptotics}

For the particular choices of $\Omega(\eta)$ we intend to consider,
analyzing the asymptotic behavior of solutions to (\ref{Y}) is
greatly simplified by the following lemma:\vspace{5mm}

\noindent {\it If the equation
\be
Y''+\Omega(\eta) Y=0
\label{Yeq}
\ee
has solutions that behave for large $\eta$ as\footnote{The relation between the powers of $\eta$ in $Y_1$ and $Y_2$ is dictated by the conservation of the
Wronskian. See also the discussion below (\ref{boundvalid}).}
\be
Y_1(\eta)\sim \eta^a + o(\eta^a),\qquad Y_2(\eta)\sim \eta^{1-a} + o(\eta^{1-a})
\label{asymp}
\ee
with $a>1/2$, then the equation
\be
{\tilde Y}''+\left(\Omega(\eta)+o(1/\eta^b)\right){\tilde Y}=0
\label{Ymod}
\ee
with $b> 2$ has solutions with the same asymptotic behavior.}\vspace{5mm}

For the sake of brevity of notation, we shall phrase the proof for the asymptotics
at $\eta=+\infty$. Take any solution to (\ref{Ymod}) and write it in the form
\be
{\tilde Y}(\eta)=\xi(\eta)Y(\eta),
\ee
where $Y(\eta)$ is the solution to (\ref{Yeq}) satisfying
\be
Y(\eta_0)={\tilde Y}(\eta_0),\qquad Y'(\eta_0)={\tilde Y}'(\eta_0)
\ee
for some positive $\eta_0$. Then $\xi(\eta)$ is a continuous function
satisfying
\be
\xi''+2\,\frac{Y'}{Y}\,\xi'+ o(1/\eta^b)\xi=0,\qquad \xi(\eta_0)=1,\qquad \xi'(\eta_0)=0,
\ee
which can be rewritten as
\be
\eta^{-2a}\left(\eta^{2a}\xi'\right)'= o(1/\eta) \xi' + o(1/\eta^b)\xi
\label{xieq}
\ee
(we have assumed that $Y(\eta)$ displays the dominant asymptotic $t^a$, rather than
the sub-dominant asymptotic $t^{1-a}$, which should be generically possible to
achieve for any $\tilde Y(\eta)$ by choosing $\eta_0$).

Let $\eta_*$ be the first $\eta>\eta_0$ for which $|\xi'(\eta)|=A \eta^{-c}$, with
$1<c<\mbox{min}(2a,b-1)$ and some positive constant $A$ (if no such $\eta_*$ exists, one can skip to (\ref{boundvalid}) without any further
considerations). Note that, by construction,
\be
|\xi'(\eta)|<A \eta^{-c},\qquad |\xi(\eta)|< 1 + \frac{A}{(c-1)\eta_0^{c-1}},\qquad\mbox{for}\hspace{5mm}\eta_0<\eta<\eta_*.
\label{bounds}
\ee
One can integrate (\ref{xieq}) from $\eta=\eta_0$ to $\eta=\eta_*$ to obtain
\be
\left|\eta_*^{2a}\xi'(\eta_*)\right|\le A \left|\,o(\eta^{2a-c})\big|_{\eta=\eta_0}^{\eta=\eta_*}\,\right|+ \left(1 +
\frac{A}{(c-1)\eta_0^{c-1}}\right)\left|\,o(\eta^{2a-b+1})\big|_{\eta=\eta_0}^{\eta=\eta_*}\,\right|,
\ee
which can be rewritten as (all the $o$-symbols are taken to be positive)
\be
\eta_*^{2a-c}\le \left(o(\eta_*^{2a-c}) + o(\eta_0^{2a-c})\right)+ \left(\frac1{A} +
\frac{1}{(c-1)\eta_0^{c-1}}\right)\left(o(\eta_*^{2a-b+1}) + o(\eta_0^{2a-b+1})\right)
\ee
or
\be
\begin{array}{l}
\dsty1\le \left(\frac{o(\eta_*^{2a-c})}{\eta_*^{2a-c}} +
\left(\frac{\eta_0}{\eta_*}\right)^{2a-c}\frac{o(\eta_0^{2a-c})}{\eta_0^{2a-c}}\right)\vspace{3mm}\\
\dsty\hspace{3.5cm}+ \left(\frac1{A} + \frac{1}{(c-1)\eta_0^{c-1}}\right)\left(\frac{o(\eta_*^{2a-b+1})}{\eta_*^{2a-c}}  +
\left(\frac{\eta_0}{\eta_*}\right)^{2a-c}\frac{o(\eta_0^{2a-b+1})}{\eta_0^{2a-c}}\right).
\end{array}
\ee
Since the right-hand side goes to 0 if $\eta_*$ and $\eta_0$ go to infinity with $\eta_*>\eta_0$ and $1<c<\mbox{min}(2a,b-1)$ (as
originally specified),
it is possible to choose $\eta_0$ and $A$ (independent of $\eta_0$) in such a way that, for all $\eta_* > \eta_0$, the
inequality is {\it not} satisfied. Then no $\eta_*$ (as defined above) exists, and instead of (\ref{bounds}), one has
\be
|\xi'(\eta)|<A \eta^{-c},\qquad\mbox{for}\hspace{5mm}\eta>\eta_0.
\label{boundvalid}
\ee
Since $c>1$, this implies that $\xi(\eta)$ goes to a (generically non-zero) constant
at infinity\footnote{Since $|\xi(\eta)-1|<A/(c-1)\eta_0^{c-1}$ with $A$ independent of $\eta_0$,
one can always avoid the vanishing of $\xi(\eta)$ at infinity by choosing a sufficiently large value of $\eta_0$.}, and the leading
asymptotics of the solution $\tilde Y(\eta)$ is the same
(up to a constant factor) as the leading asymptotics of $Y(\eta)$.

The existence and asymptotic behavior of the ``subdominant'' ($\eta^{1-a}$)
solution can be inferred from the following consideration. The leading
solution of (\ref{Ymod}) that we have just analyzed does not oscillate as
$\eta$ goes to $+\infty$. Hence, the equation (\ref{Ymod}) itself is non-oscillatory at $\eta=+\infty$
(see \cite{zettl} for the relevant definitions and properties).
Thus, it must have a solution (unique up to a constant) that grows slower than $\eta^a$ at infinity
(the ``principal'' solution), and by conservation of the Wronskian \cite{zettl},
this solution is
\be
\tilde Y_2(\eta) \sim \tilde Y_1(\eta)\int\limits_{\eta}^{+\infty}\frac{d\tilde\eta}{\tilde Y_1^2(\tilde\eta)},
\label{Wr}
\ee
with $\tilde Y_1$ being a dominant (``non-principal'') solution to (\ref{Ymod}).  Since the
asymptotic behavior of $\tilde Y_1$ has been established as
\be
\tilde Y_1(\eta)\sim \eta^a + o(\eta^a),
\ee
together with (\ref{Wr}), this gives
\be
\tilde Y_2(\eta)\sim \eta^{1-a} + o(\eta^{1-a}).
\ee
This concludes the proof of the lemma.

\subsection{Specific cases}

Armed with the asymptotic behavior lemma, we can analyze the singular limit
for specific pp-waves of interest. Namely, we shall assume that $\Omega$
of (\ref{F}) behaves as
\be
\Omega(\eta)= \frac{k_{\pm}}{\eta^2} + O(1/\eta^{b})
\label{Omega}
\ee
for $\eta$ going to $\pm\infty$ (respectively), for some constants $k_{\pm}$
and $b>2$.

We shall be interested in the case $0\le \lambda k_\pm < 1/4$. For these values, on can
transform to the so-called Rosen co-ordinates, for which the $\eps\to 0$ limit of the metric takes the form
\be
ds^2=-dX^+dX^-+\left(X^+\right)^{2(1-a_{\pm})}\,dX^2,
\label{rosen}
\ee
with $a_{\pm}$ corresponding to $X^+>0$ and $X^+<0$ respectively,
\be
a_\pm=\frac12+\sqrt{\frac14-\lambda k_\pm}.
\ee
The metric (\ref{rosen}) is somewhat reminiscent of Friedman cosmologies (and arises from those in the Penrose limit \cite{limit}). The case $0< \lambda k_\pm < 1/4$ has
been termed ``null-cosmology''. The case
$k_\pm=0$ refers to well-localized pp-waves, and, in the $\eps\to 0$ limit, one
is left with Minkowski space everywhere away from $X^+=0$, and a strong singularity
at $X^+=0$. The case of the ``light-like reflector plane'' described in \cite{g-res}
falls precisely into this latter category.

By the asymptotic behavior lemma, the equation
\be
Y''+\lambda \Omega(\eta) Y=0
\label{YYY}
\ee
with $\Omega$ of the form (\ref{Omega}) will have two solutions going as
\be
Y_{1-}(\eta)=|\eta|^{a_-}+o(|\eta|^{a_-}),\qquad Y_{2-}(\eta)=|\eta|^{1-a_-}+o(|\eta|^{1-a_-})
\ee
for $\eta$ approaching $-\infty$, and two solutions going as
\be
Y_{1+}(\eta)=\eta^{a_+}+o(\eta^{a_+}),\qquad Y_{2+}(\eta)=\eta^{1-a_+}+o(\eta^{1-a_+})
\ee
for $\eta$ approaching $+\infty$. The two sets of solutions are of course related:
\be
\left(\begin{array}{l}Y_{1-}\vspace{1mm}\\Y_{2-}\end{array}\right)=C(\lambda)
\left(\begin{array}{l}Y_{1+}\vspace{1mm}\\Y_{2+}\end{array}\right),
\label{C}
\ee
where $C(\lambda)$ is a $2\times 2$-matrix.

Since we are interested in the evolution across the singular point $t = 0$, we choose an initial time $t_1 < 0$ and final time $t_2 > 0$.  By (\ref{comp}),
\be
\mathcal{C}(t_1,t_2) = \eps\,\frac{Y_{1-}(t_1/\eps)Y_{2-}(t_2/\eps) - Y_{1-}(t_2/\eps)Y_{2-}(t_1/\eps)}{W_{\eta}[Y_{1-},Y_{2-}]},
\ee
where the Wronskian $W_\eta$ is evaluated with respect to $\eta$.
Using the asymptotic expansion of solutions in the past and future as given above together with (\ref{C}) and $W_{\eta}[Y_{1-},Y_{2-}]=1-2a_-$, we obtain
\be
\begin{array}{l}
\dsty{\cal C}(t_1,t_2)= \frac{C_{11}(\lambda)}{2a_--1}\,|t_1|^{1-a_-}t_2^{a_+}\eps^{a_--a_+}+\frac{C_{12}(\lambda)}{2a_--1}\,|t_1|^{1-a_-}t_2^{1-a_+}\eps^{a_-+a_+-1}\vspace{3mm}\\
\dsty\hspace{4.5cm}-\frac{C_{21}(\lambda)}{2a_--1}\,|t_1|^{a_-}t_2^{a_+}\eps^{1-a_--a_+}-\frac{C_{22}(\lambda)}{2a_--1}\,|t_1|^{a_-}t_2^{1-a_+}\eps^{a_+-a_-}.
\end{array}
\label{complimit}
\ee
We demand the $\eps\to 0$ limit of (\ref{complimit}) to exist for all fixed $t_1$ and $t_2$.  Simple inspection shows that the leading power of $\eps$, i.e., the one that is the most negative, occurs in the term proportional to $C_{21}(\lambda)$, and that the associated power of $\eps$ is always negative.  Thus, if the
$\eps\to 0$ limit exists, we must have
\be
C_{21}(\lambda)=0.
\label{discrete}
\ee
The power of $\eps$ in the term proportional to $C_{12}$ is always positive,
which makes it simply vanish in the limit. Furthermore, if $a_+\ne a_-$ (i.e., $k_+\ne k_-$), one must have either
\be
C_{11}(\lambda)=0
\label{empty1}
\ee
or
\be
C_{22}(\lambda)=0
\label{empty2}
\ee
(depending on the sign of $a_+ - a_-$).

Under generic conditions, (\ref{discrete}) will leave only a discrete set
of allowed values for $\lambda$.  Indeed, (\ref{discrete}) merely states that the solution to (\ref{YYY}) subdominant
at $\eta=-\infty$ does not receive any admixture of the dominant solution at $\eta=+\infty$, i.e. that the solution is subdominant at both $\pm\infty$. This is essentially a Sturm-Liouville problem, and the appearance of a discrete spectrum\footnote{When $k_\pm = 0$, the subdominant solution to (\ref{YYY}) is bounded at infinity. One can use the integral representation for such bounded solutions \cite{AM} to show that the two bounded solutions at $\pm\infty$ depend analytically on $\lambda$.  The Wronskian of these two solutions (proportional to $C_{21}(\lambda)$) is then a non-identically-vanishing analytic function of $\lambda$.  Hence its zeros, which are the solutions to (\ref{discrete}), are discrete.  For $k_\pm \neq 0$, some further analysis is needed to prove rigorously the discreteness of the spectrum for $\lambda$.} should be hardly surprising. A particular exactly solvable example for this discrete spectrum (there called ``light-like reflector plane'') has
been given in \cite{g-res}.

Imposing additionally (\ref{empty1}) or (\ref{empty2}) would make the determinant of $C(\lambda)$ vanish, in contradiction with the conservation of the Wronskian, and hence (\ref{YYY}). We thus conclude that the $\eps\to 0$ limit will exist for the solutions of the wave equation (in the class of pp-wave backgrounds we have been considering) if and only if $\lambda$ belongs to a {\it discrete spectrum} of values, and the leading $(1/X^+)^2$ {\it asymptotics}
of the pp-wave profile are {\it the same} on the both sides of the singularity.

\section{Discreteness and matching conditions}

In this note, we have considered the wave equation in families of pp-wave geometries
developing strong scale-invariant singularities in certain limits. (The requirement
of scale invariance is quite constraining, but one may expect that some of our results
will extend to more general cases, since the presence of finite scale parameters
is not likely to affect the dynamics in an infinitesimally small singular region.)

We have observed that the singular limit of the solutions to the wave equation with initial data specified away from the singularity exists only if:
\begin{itemize}
\item the absolute normalization of the pp-wave profile lies in a discrete spectrum (dependent on the specific way the singularity
    is resolved);
\item the leading $(1/X^+)^2$ asymptotics of the pp-wave profile are the same before and after the singularity.
\end{itemize}
Even though these requirements are tremendously constraining, their predictive power is diminished in our present setting by the
complete arbitrariness
of the pp-wave profiles. One could speculate however, that if the time dependence
of the pp-wave is governed by further specifications (as it might be, for example,
if the pp-wave arises as a Penrose limit of some other geometry; see also footnote \ref{penrose}), the discrete features
we observe will relate the structure of the singularity to some parameters
of the space-time geometry away from the singular region.

\section*{Acknowledgments}

O.E. would like to thank Ben Craps and Frederik De Roo for collaboration on closely related subjects, and Matthias Blau and Martin O'Loughlin for valuable comments. 
The research of O.E. has been
supported in part by the Belgian Federal Science Policy Office through the Interuniversity Attraction Pole IAP VI/11, by the European
Commission FP6 RTN programme MRTN-CT-2004-005104 and by FWO-Vlaanderen through project G.0428.06. T.N. would like to thank Barry Simon for helpful discussions.

\end{document}